\documentclass[10pt,showpacs,amsmath,amssymb,pre,floatfix,notitlepage,superscriptaddress]{revtex4-1}

\usepackage{graphicx}
\usepackage{epstopdf}
\usepackage{dcolumn}
\usepackage{bm}
\usepackage{color}
\usepackage{forloop}
\usepackage{amsfonts}
\usepackage{mathtools}
\usepackage[normalem]{ulem} 
\usepackage[colorlinks=true,
            linkcolor=black,
            urlcolor=blue,
            citecolor=blue]{hyperref}

\definecolor{blue}{RGB}{0, 0, 150}
\definecolor{green}{RGB}{0,150,0}
\definecolor{red}{RGB}{200, 0, 0}
\definecolor{black}{RGB}{0, 0, 0}

\definecolor{ao}{rgb}{0.0, 0.5, 0.0}

\begin{document} 
\title{A stochastic and dynamical view of pluripotency in mouse embryonic stem cells}
\author{Yen Ting Lin}
\affiliation{T-6 and Center for Nonlinear Studies, Los Alamos National Laboratory, Los Alamos, NM 87545, USA}
\affiliation{School of Physics and Astronomy, The University of Manchester, Manchester, M13 9PL, UK} 
\author{Peter G. Hufton}
\affiliation{School of Physics and Astronomy, The University of Manchester, Manchester, M13 9PL, UK} 
\author{Esther J.~Lee}
\affiliation{Department of Bioengineering, Rice University, Houston, TX 77005, USA}
\author{Davit A. Potoyan}
\affiliation{Department of Chemistry, Iowa State University, Ames, IA 50011, USA}
\date{\today}

\begin{abstract}

Pluripotent embryonic stem cells are of paramount importance for biomedical research thanks to their innate ability for self-renewal and differentiation into all major cell lines. The fateful decision to exit or remain in the pluripotent state is regulated by complex genetic regulatory network. Latest advances in transcriptomics have made it possible to infer basic topologies of pluripotency governing networks. The inferred network topologies, however, only encode boolean information while remaining silent about the roles of dynamics and molecular noise in gene expression. These features are widely considered essential for functional decision making. Herein we developed a framework for extending the boolean level networks into models accounting for individual genetic switches and promoter architecture which allows mechanistic interrogation of the roles of molecular noise, external signaling, and network topology. We demonstrate the pluripotent state of the network to be a broad attractor which is robust to variations of gene expression. Dynamics of exiting the pluripotent state, on the other hand, is significantly influenced by the molecular noise originating from genetic switching events which makes cells more responsive to extracellular signals. Lastly we show that steady state probability landscape can be significantly remodeled by global gene switching rates alone which can be taken as a proxy for how global epigenetic modifications exert control over stability of pluripotent states.

\end{abstract}
\maketitle

\section{Introduction} 

Embryonic stem cells derived from mammalian blastocyst are pluripotent: they show an indefinite capacity for self-renewal and the ability to differentiate into every cell type constituting an adult organism~\cite{evans2011discovering,murry2008differentiation,martello2014nature}. The development of healthy tissues hinges on the ability of these pluripotent stem cells to make critical decisions determining when and into which kind of cells to differentiate in response to their environment. Fates of embryonic cells are therefore decided through sophisticated biological computations orchestrated by a vast regulatory network consisting of genetic, epigenetic and signaling layers~\cite{martello2014nature}. 

The dynamics of genetic expressions take place in molecular environments and are subject to intrinsic noise due to the discrete nature of the molecular copy numbers  and to extrinsic noise from extracellular environment \cite{van1992stochastic}. 
Thus, while at the level of the organism development is often predictable with a well-defined order of events, at the level of single cells fate determination is fundamentally stochastic \cite{symmons2016s,balazsi2011cellular}. 
An outstanding question involves understanding to what extent the inherent molecular stochasticity in embryonic cells is suppressed or exploited for functional purposes. 
Studies probing transcription in single cells have uncovered high variabilities of gene expression in embryonic stem cell populations \cite{singer2014dynamic, kumar2014deconstructing,canham2010functional}. 
Several hypotheses about the functional roles of stochasticity in embryonic stem cells have been put forward. 
Observations of dynamic and heterogeneous expression patterns of core transcription factors such as Nanog, Oct4 and Sox2 have promoted the view that stochastic excursions of these factors are governing the stability of pluripotent states~\cite{kalmar2009regulated,masui2007pluripotency}. 
A different hypothesis claims transcriptional noise to be advantageous by facilitating the exploration of the space of a gene network such that, at any instant, a subpopulation of cells is optimally primed to be responsive to differentiation signals \cite{macarthur2013statistical}. 
The heterogeneity of populations of pluripotent cells has also raised some concerns that pluripotency is ill-defined on a single-cell level~\cite{macarthur2013statistical} and instead should be viewed as a statistical property of the macroscopic state emerging at the level of an ensemble of cells. 
A comprehensive physical picture of pluripotency at the single-cell level therefore still remains unclear.  

\begin{figure*}[!th]
\centering
\includegraphics[width=0.95\textwidth]{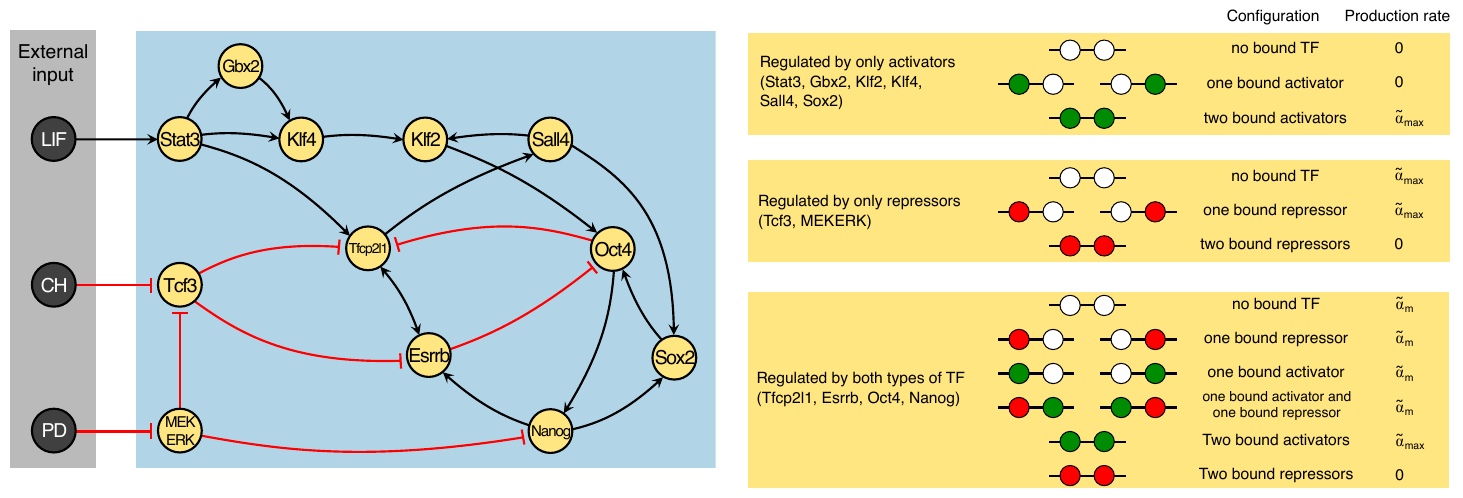}
\caption{The left panel shows a schematic diagram of the network topology, reproduced from Dunn {\em et al.}~\cite{dunn2014defining}. Each node corresponds to a given gene and their placement from left to right is chosen to indicate a general trend of downstreamness from the inputs. In our mechanistic model, each gene produces a unique transcription factor at a given rate, depending on the binding state of its promoter sites. These transcription factors go on to bind and activate (black arrow) or repress (red bar) other genes. The three nodes on left correspond to extra-cellular signals, which are either absent or present.
The right panel shows our assumed molecular logic of transcriptional regulation. Shown are the outcomes of combinatorial binding of activators and repressors to promoter sites. Depending on the configuration of the promoter site, transcription factors are produced with rates 0, $\tilde\alpha_m$, and $\tilde\alpha_\text{max}$, modeling the effects of cooperative binding. 
}
 \label{fig:schematic}
\end{figure*}

In this regard, the roles of modeling and computational approaches are seen as especially important for bridging the gap between our understanding of molecular dynamics of regulatory networks and phenotypic outcomes. 
The rapid accumulation of data of gene expression through single-cell experiments has stimulated the techniques of statistical inference and computational systems biology for mapping embryonic stem cell networks \cite{feigelman2016analysis,xu2014construction,semrau2015studying}. In vitro studies of mouse embryonic stem cells (mESC) in different culture conditions, in particular, have become an ideal model system for exploring mechanistic issues surrounding pluripotency and lineage commitment in mammals~\cite{semrau2015studying}. In a {\it tour de force} study of mESC by Dunn {\em et al.}~\cite{dunn2014defining}, regulatory relationships between transcription factors were uncovered through analysis of pairwise correlations in gene expression. Using known experimental data as constraints, a network topology with a minimal set of genes was derived with a high degree of predictive power with respect to perturbations of the network. 

Network topologies, however, remain silent about the roles of molecular noise and dynamics in stem cell differentiation governed by stochastic biochemical reactions. Furthermore, in order to validate that the inferred network reflects true microscopic reality of cell and is not a result of overfitting, one has to ultimately test the results using mass-action-based kinetics which integrate relevant molecular factors. The key challenge lies in finding the adequate resolution for the network which is able to be predictive and does not pose insurmountable computational burden.

In the present work, we outline a framework for extending boolean resolution networks into stochastic and dynamical models---with microscopic resolution of promoter architecture and individual gene switching events.  The computational scheme utilizes static boolean information about the network topology and uses novel analytical model reduction to increase the computational efficiency, allowing for extensive searches in the space of microscopic reaction rates. This framework is successfully applied to the network topology inferred by Dunn {\em et al.}~\cite{dunn2014defining} in order to build a mass action based stochastic dynamic model which is capable of describing both the discrete states of all the genes and the populations of transcription factors. Starting from minimal assumptions about the rates of various reactions, we find a few distinct gene swithcing regimes where a remarkable agreement with the experimental gene expression profiles is achieved for all combinations of experimentally controlled external signals. We show that gene expression levels in complex regulatory networks are not a unique function of gene switching rates which cautions against over-interpreting boolean level networks and suggests strategies of inference which utilize higher moments in distribution of transcription factors. Using single cell experiments which have probed expression of pluripotency factors~\cite{singer2014dynamic, kumar2014deconstructing} we are able to argue that gene switching in pluripotent states happens primarily on the intermediate scale relative to the reactions of creation and degradation (dilution). This regime better agrees with the diverse set of experiments available and provides explanation for the multimodality in distributions of transcription factors and burst like expression dynamics for some genes.  

In the second half of the paper, armed with a predictive and physically motivated model of pluripotency network, we explore the dynamics of lineage commitment driven by withdrawal of various well documented signals (LIF, 2i) for maintaining the na\"\i ve state of pluripotency. We find a number of non-trivial consequences of molecular noise and gene switching dynamics. In particular we show that intermediate gene switching regime generates higher sensitivity for the network when responding to external signals. 

\section{Framework for deriving microscopic resolution networks from experimentally inferred network topologies}

\subsection{Molecular logic of transcriptional regulation in ESC}

Here we outline the basic steps of constructing the microscopic genetic regulatory networks from the Boolean topology of genes inferred from experiments~\cite{dunn2014defining}. 
The network topology (Fig.~\ref{fig:schematic}) contains static information about the types of interactions between pairs of genes. 
The interactions are classified as being either repressing or activating. To study the dynamics of complex genetic networks, one has to extend the Boolean level description to account for the molecular logic of gene regulation. 
This molecular logic specifies the precise relation between the binding of transcription factors to genomic sites and the regulatory outcome in terms of gene activation, repression, etc. 
In the case where the same sites can be bound to different transcription factors, the combinatorial nature of regulation can give rise to ambiguity in molecular logic. 
Such ambiguities can be resolved either by directly inferring regulatory logic or by simulating different combinatorial possibilities until sufficient agreement with experiments is reached. 
Once molecular logic is defined the topologic level description can be extended into a set of biochemical reactions.

The reactions in the present model include {\bf (i)} the binding/unbinding of transcription factors to promoter sites of genes and {\bf (ii)} the creation/degradation of populations of transcription factors. 
We adopt the network of mESCs inferred by Dunn {\em et al.}~\cite{dunn2014defining}. 
There are $N=12$ genes in the network and we assign each gene $G_i$ to a discrete variable which describes the configuration of its promoter sites. 
The values of $G_i$ determine if the transcriptional activity is in the activated, repressed or neutral state. 
The $n$ promoter sites of a particular gene allow binding of at most $n$ transcription factors.
Since a large majority of eukaryotic transcription factors in this network bind as dimers~\cite{ferraris2011combinatorial} we set $n=2$ for all genes \footnote{We have also tested $n=3,4,5$ showing qualitatively similar results to $n=2$}. The binding reactions are 
\begin{equation} \label{eq:promoterDynamics}
\begin{aligned}
\text{First TF binding:~~~}  \phantom{P_j}G_i+P_j & \xrightleftharpoons[\tilde{k}_\text{off}]{2\times\tilde{k}_\text{on} N_{P_j}\,} G_i P_j,\\
\text{Second TF binding:~~~}   G_i P_j +P_k & \xrightleftharpoons[\tilde{k}_\text{off}]{\,\tilde{k}_\text{on} N_{P_k}\,} G_i P_{jk}. \\
\end{aligned}
\end{equation}
Notice that we have adopted the uncooperative binding mechanism  which assumes the transcription factors bind to any of the unbound promoter sites independently with a constant rate $\tilde{k}_\text{on}$. Other cooperative binding mechanisms can be easily built into this framework. Each gene codes for its own transcription factor $P_i$ where indices $i=1,...,N$ label both genes and the corresponding transcription factors.  These promoters can act as repressors or activators of other genes: this information is contained in the experimentally-inferred topology.  The molecular logic describing how the combinatorial binding of activators and repressors regulate each gene, however, is an assumption in the model.

When the rate of transcription is non-zero, genes produce copies of their respective transcription factors. We use a one step model of transcription~\cite{bressloff2014stochastic} whereby transcription factors are produced via a unimolecular step. This approximation coarse grains several sequential intermediate reactions, such as mRNA production, into one effective step~\cite{kepler2001stochasticity,hornos2005self}. This approximation has become popular in models of protein production~\cite{lin2016bursting, lin2016gene, hufton2016intrinsic}. All of the transcription factors are assumed to have finite lifetimes set by the rate of degradation ensuring the existence of stable steady states with finite number of molecules. The reactions are
\begin{equation}\label{eq:TFDynamics}
\begin{aligned}
\text{TF production:~~~}       G_i P_{jk} & \xrightarrow{\tilde\alpha_{jk}} G_i P_{jk} +P_i,\\
\text{TF degredation:~~~}       \phantom{G_{jk}}P_i & \xrightarrow{\tilde\gamma} \varnothing. \\
\end{aligned}
\end{equation}
We note that one can adopt a view with higher resolution of the network depending on the available experimental data and the type of question posed. For instance one may consider explicitly including the steps of cell cycle regulation, different epigenetic states, binding of non-coding RNAs etc. For simplicity and illustrative purpose, we consider the most simplified dynamical model which describes only the promoter configurations and the population dynamics of the transcription factors---this model aims to use the optimal resolution for capturing trends in gene expression while remaining feasible for efficient stochastic simulations capturing the variability in an ensemble of cells. 
After converting the network topology into a higher resolution network of biochemical reactions, our next goal is to exhaustively sample a vast space of parameter space to identify the optimal parameter regime with which the model best reproduces all of the experimental constraints.

\begin{figure*}[!t]
\centering
\includegraphics[width=0.95\textwidth]{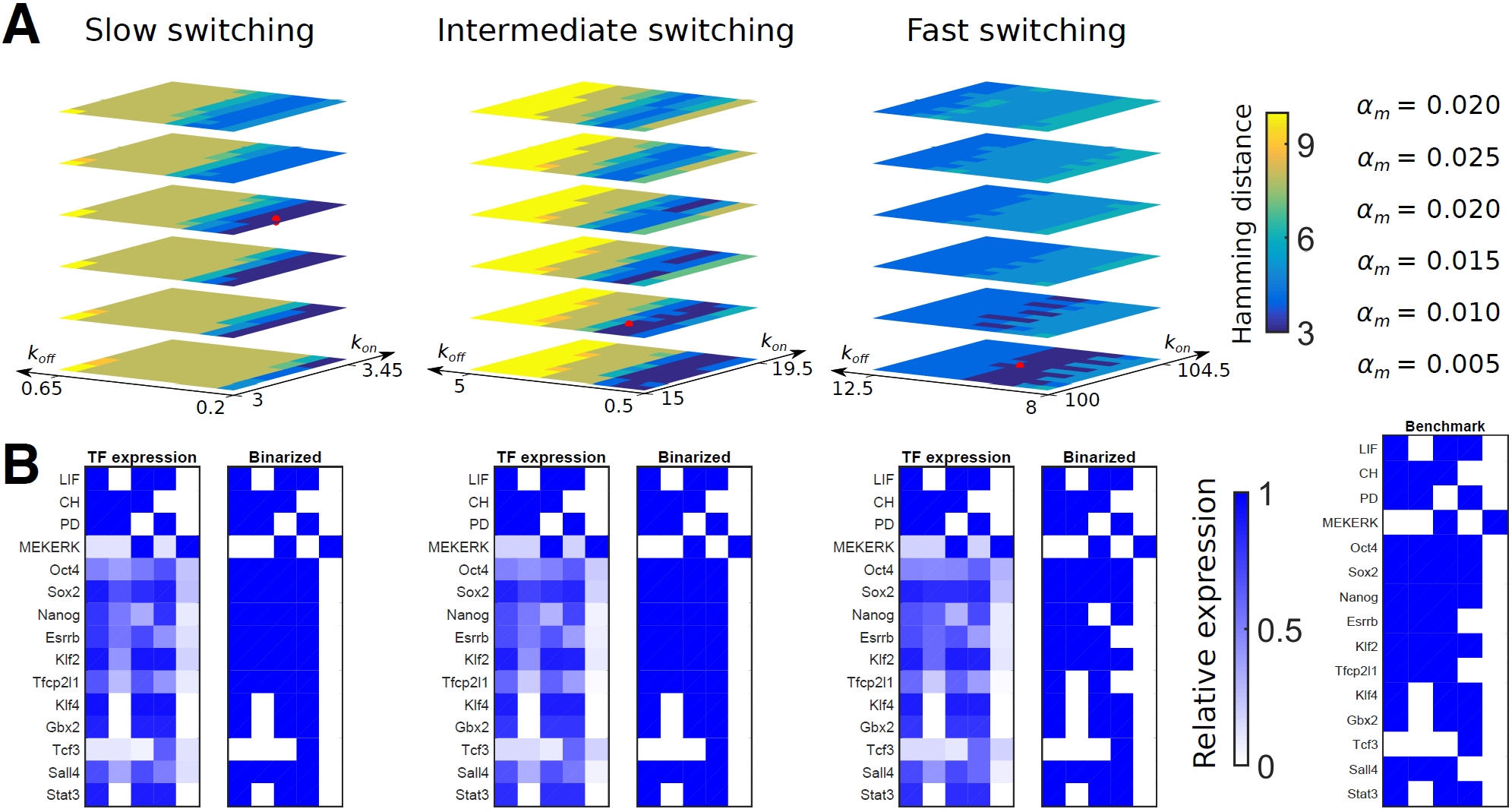}
 \caption{PDMP stochastic simulations identify three genetic switching regimes consistent with experimental data. When switching is slow, intermediate, or fast we find certain parameters which closely match the experimental results obtained by Dunn {\em et al.}~\cite{dunn2014defining}. The consistency of our model with experimental results is measured using a Hamming distance---a measure where one counts the number of discrepancies between the binary expression of each TF for both cases. (A) Shown are the identified regions in parameter regimes that minimize Hamming distance. There are three free parameters: the binding rate $k_\text{on}$, unbinding rate $k_\text{off}$, and basal transcription rate $\alpha_m$. For slow switching, the parameters are $k_\text{on}=3.2$, $k_\text{off}=0.2$, $\alpha_m=0.02$; for intermediate switching, $k_\text{on}=16$, $k_\text{off}=1.5$, $\alpha_m=0.01$; for fast switching, $k_\text{on}=102$, $k_\text{off}=10$, $\alpha_m=0.005$. The selected parameter sets are presented as red dots in the landscapes in the upper panel. (B) Comparison of computed and discretized gene expression profiles with that of the experiments (Benchmark panel). }
\label{fig:landscape}
\end{figure*}

\subsection{Multi-scale simulation of complex genetic networks}
Biochemical reactions in gene networks are of fundamentally probabilistic nature. 
The most rigorous way to simulate a network of reactions is by numerically solving the master equation which accounts for all possible states of the network down to the level of single molecules \cite{van1992stochastic,bressloff2014stochastic}. 
However in high dimensions, i.e., when the number of species is large, this approach is not computationally efficient.
Instead, kinetic Monte Carlo algorithms are the most straightforward way to generate sample paths of the stochastic process. 
Fully individual-based models, however, still suffer from a steep scaling of computational time with the number of components.
This fact renders them inefficient for simulating large gene networks, especially when it comes to scanning or exploring the parameter space. 

A wide range of approximate schemes have been employed to simulate large scale gene regulatory networks. 
Most conventional approximations so far have been the size-expansion methods which are known for being problematic when the molecular noise induced by genetic switching becomes non-negligible. 
On the other hand, for embryonic stem cell networks it is essential to account for the stochastic nature of genetic switches which give rise to multiple attractor states corresponding to various phenotypes. 
A network with $N$ genes can theoretically have $\sim 2^N$ attractors. 
Even if populations of all the other species are present in large quantities, the stochastic fluctuations caused by the genetic switches (due to stochastic binding-unbinding events of the transcriptional factors) between ON and OFF states cannot be ignored, unless the switching is operating in the extremely fast limit compared to any other reactions, known as the ``adiabatic regime" \cite{potoyan2015dichotomous,walczak2005absolute,sasai2003stochastic}. 
In the other cases---the non-adiabatic regime---gene switching can completely dominate the dynamics in the network \cite{potoyan2015dichotomous,feng2010adiabatic}. 
Eukaryotic gene regulatory networks are often found in the intermediate regime where gene switching events are dynamically interwoven with the rest of the reactions in the network and cannot be ignored \cite{lenstra2016transcription, symmons2016s}. 
Single-cell studies of mESC in particular have shown bursty behavior in gene expression with sudden jumps in levels of proteins resulting in multi-modal distributions of core transcription factors~\cite{singer2014dynamic}. 

Many of the early computational models of embryonic stem cells have focused on small fragments of pluripotency network, typically involving bistable switches~\cite{chickarmane2006transcriptional, kalmar2009regulated}. These early pioneering studies have yielded many insights on stochastic decision making in regards to fate determination and self-renewal.  A few studied have looked at larger portions of the regulatory network~\cite{wang2011quantifying, li2013quantifying, zhang2014stem, feng2012new} but at the cost of making near-adiabatic switching approximations thereby ignoring explicit stochastic treatment of genetic switching dynamics. Lastly, with very few exceptions, the kinetic parameters in those models were not throughly explored or data-informed and had to be picked relying on physical intuition alone. All of these limitations have been completely overcome in the present work. A series of recent studies~\cite{zeiser2008simulation,zeiser2010autocatalytic,potoyan2015dichotomous, lin2016bursting, lin2016gene, hufton2016intrinsic} on gene expression dynamics have developed a novel computational framework utilizing piecewise-deterministic Markov process (PDMP) \cite{davis1984piecewise} as approximations to the fully individual-based model. 
This approach treats genetic switching events exactly, while assuming noise due to the finite nature of populations of transcription factors to be small in comparison. 
As will soon be seen, the assumptions underlying the PDMP approach turn out to be sound as we go on to obtain a nearly perfect quantitative agreement with full blown kinetic Monte Carlo schemes even for the case of the complex networks of ESC operating in the intermediate gene switching regime. 

The PDMP simulations carried out in the present study show nearly $\mathcal{O} (10^3)$ fold faster generating stochastic trajectories compared to conventional individual-based kinetic Monte Carlo techniques (the algorithm for generating PDMP sample paths is provided in the Supplementary Information). This rigorous and rapid sampling of gene switching events has not only allowed us to investigate the stochastic dynamics of the regulatory network at a longer timescale compared to conventional kinetic Monte Carlo methods, but also enabled us to explore a vast parameter space efficiently. We have used the obtained information to derive microscopic resolution models of ESC. 
The mathematical details of the model described in greater detail can be found in the SI.

\section{Results}

\subsection{In the pluripotent state the mean levels of gene expression are a hardwired feature of network architecture whereas heterogeneity is shaped significantly by stochastic dynamics of genetic switches.}

In this section we first identify certain parameters which provide results which are consistent experiment. Following this the model becomes predictive about dynamical features which are not contained in the topology. 

Having reduced the individual-based model to a computationally tractable PDMP, we are at a point where we can tune the model parameters to match experimental data.
The experimental data used for constraining the rate coefficients are the binarized gene expression levels of transcription factors under well-defined culture conditions consisting of different combinations of leukemia inhibitory factor (LIF), glycogen synthase kinase 3 (CH), and mitogen-activated protein kinase (PD). 
Identical culture conditions were used by Dunn {\em et al.} \cite{dunn2014defining} to infer the original Boolean topology are LIF, 2i (same as PD+CH), LIF+2i, LIF+PD, LIF+CH. 
We employ the Hamming distance (the number of discrepancies between the simulated and experimental profiles) as a cost-function for optimization. 
The Hamming distance is minimized through multiple rounds of simulations where we vary the three free parameters: the rates of promoter binding $k_\text{on}$ and dissociation $k_\text{off}$, and the basal transcription rate $\alpha_m$. 
Through this procedure we find that, in certain parameter regimes, our assumed molecular logic closely reproduces the gene expression profiles from experiments for different combinations of external signals. 
The dimeric binding---modeled in our framework as two binding sites ($n=2$)---of TF to promoter sites in particular leads to a more sensitive switching and emerges as an important feature of the network which we show to be advantageous for reproducing experimental data \footnote{Our analysis showing $n\ge 2$ significantly reduces the Hamming distance, compared to a single-site binding $n=1$.}. 
We find localized parameter sets in three distinct regimes of gene switching: slow, intermediate and fast; all of which successfully reproduce the experimental gene expression patterns (Fig.~\ref{fig:landscape}) corresponding to pluripotent and lineage committed cells. 
The fact that the rate parameters of the network occupy finite regions and cover different regimes suggests that distinct gene expression profiles can tolerate fluctuations in reaction rates. Such rate fluctuations, reflecting the effect of extrinsic noise, are inevitable in dynamic cellular environments of embryonic cells which experience frequent epigenetic and extracellular perturbations~\cite{torres2014transcription,semrau2015studying,ochiai2014stochastic}. In a way changes in gene switching rates can be seen as a proxy for how global epigenetic changes govern the rates of transcription factor binding to target genomic regions.

\begin{figure*}[!t]
\centering
\includegraphics[width=0.95\textwidth]{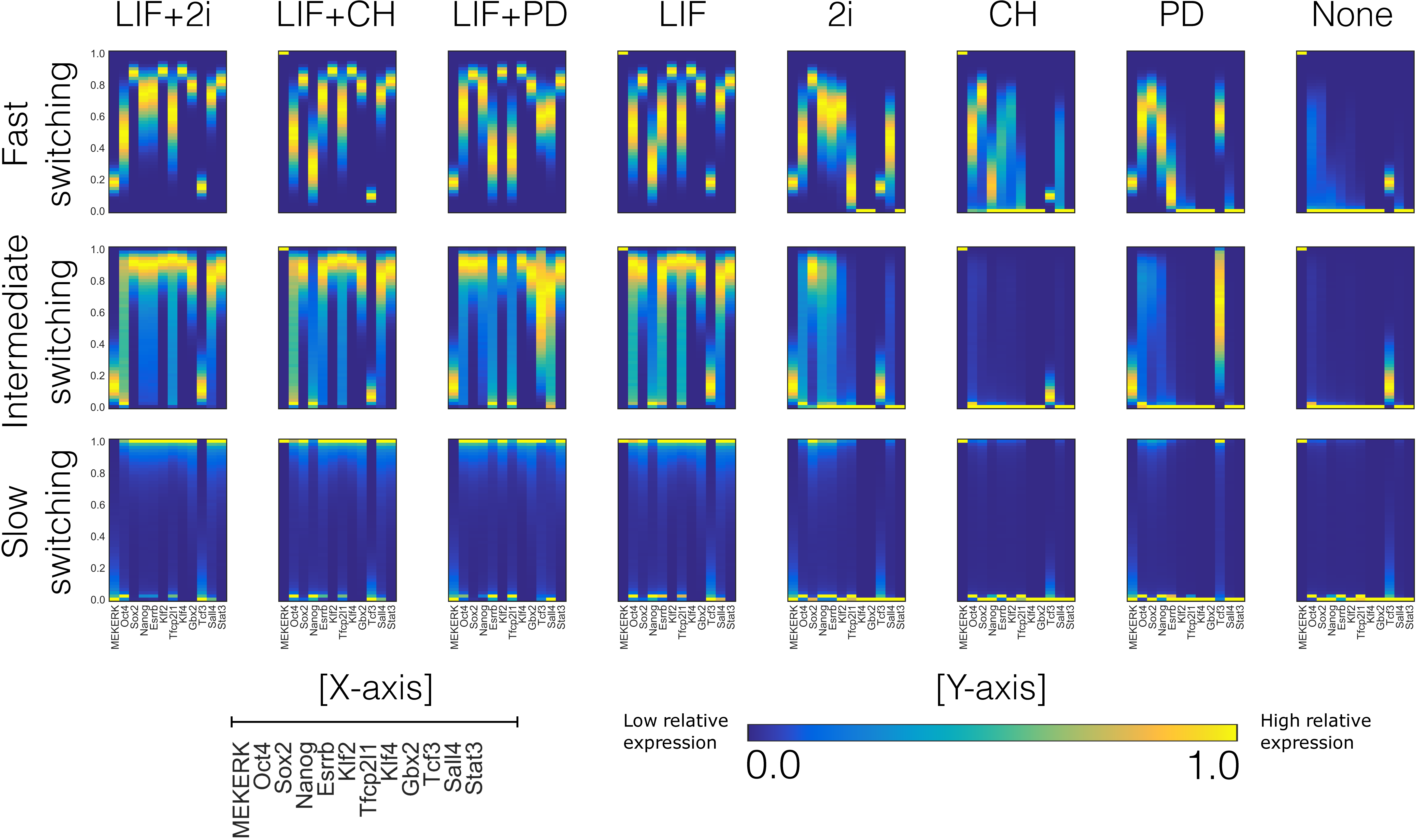}
 \caption{Gene expression profiles of pluripotency factors predicted by PDMP simulations. Each column corresponds to different external inputs, and each row corresponds to regimes of slow, intermediate and fast gene switching. More than $10^5$ sample paths were used for generating each condition.}
 \label{fig:TF expression}
\end{figure*}

From the methodology point of view the absence of a unique regime of rates implies the following: inferring networks using only mean levels of gene expression (as is done for Boolean networks) may lead to the loss of valuable information contained in higher moments of distribution. Thus new approaches of inference need to be developed in order to account for broad distributions of transcription factors. 
For this reason, we look beyond comparisons of mean expression levels and turn to comparing stationary distributions of transcription factors observed in experiments with the computationally generated distributions in three chosen parameter regimes identified in Fig.~\ref{fig:landscape}. The pluripotent state of mESCs has been the subject of intense investigations by nucleic acid-based single-cell techniques such as RNA-seq, sm-FISH, qPCR and there is now extensive data on the steady state distributions of RNA and transcription factors maintained under pluripotency-favoring culture conditions \cite{kumar2014deconstructing, singer2014dynamic, klein2015droplet}. 
These experiments have revealed a heterogeneous nature of gene expression with many core pluripotency factors, such as Nanog, having long tailed or bimodal distributions. 
Additionally, the single-cell stochastic trajectories of TFs have shown sharp, bursty transitions implying infrequent genetic switching events \cite{singer2014dynamic}. 
These experimental observations are more consistent with the intermediate regime of genetic switching as seen from Fig.~\ref{fig:TF expression}, as in the case of slow switching nearly all of the transcription factors express bimodality and in the case of fast switching the expressions of all the factors are narrow and unimodal. 

\begin{figure*}[!t]
\centering
\includegraphics[width=0.98\textwidth]{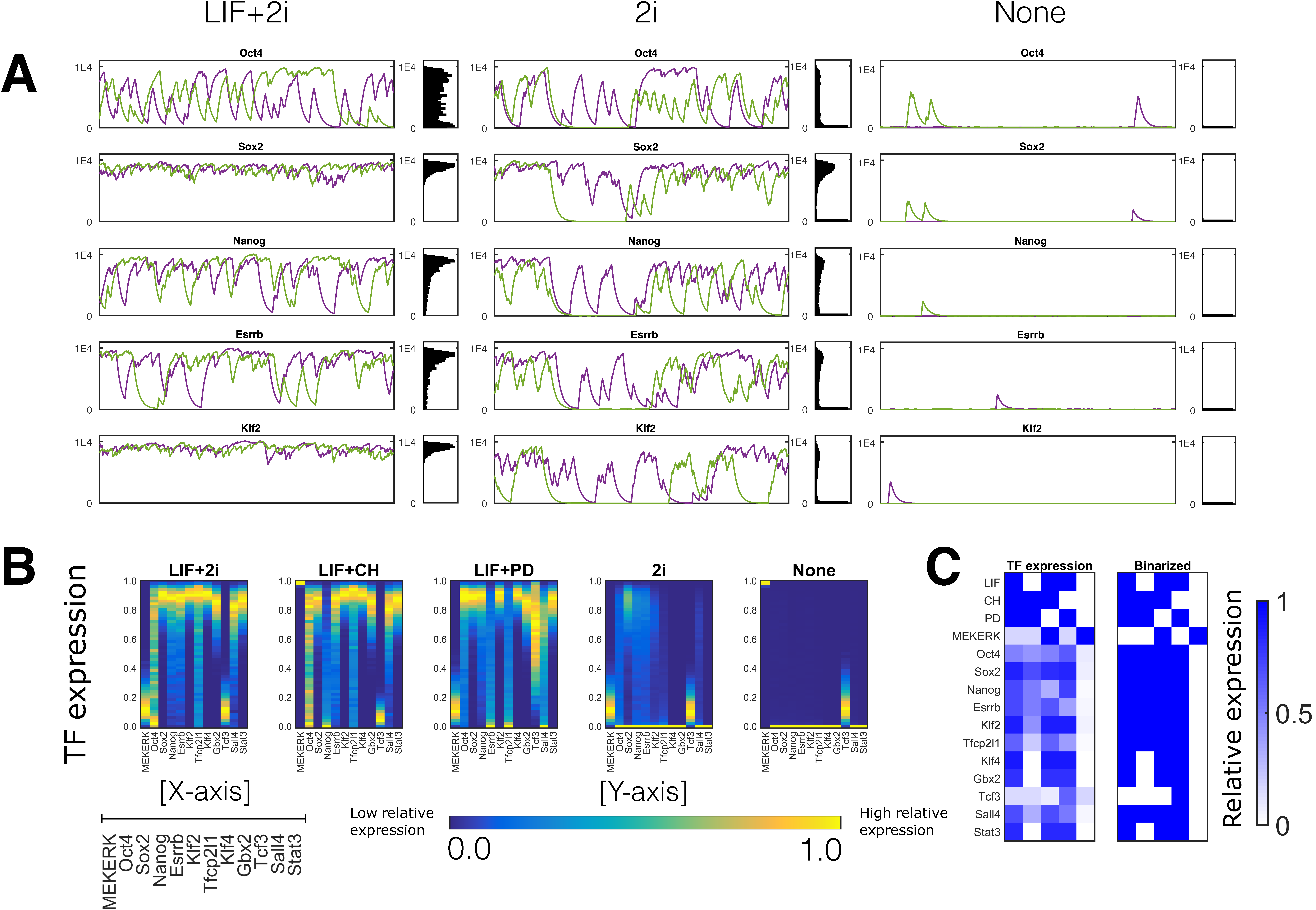}
 \caption{Gene expression profiles of pluripotency factors predicted by individual-based simulations. 
The intermediate switching regime is chosen to be presented as it is the regime which best captures the experimentally measured distributions~\cite{kumar2014deconstructing, singer2014dynamic}. (A) Shown are 100 representative trajectories and full distributions of select few transcription factors under three different conditions generated by individual based simulations. (B) Gene expression profile showing the near quantitative agreement with results of PDMP simulations shown on Fig. 3 (C) Comparison of individual based model with experimental data from Dunn et al~\cite{dunn2014defining} }
 \label{fig:IBM}
\end{figure*}

Although the PDMP approach accurately captures the effects of genetic switching, it assumes demographic noise arising from finite populations to be negligible. To test the validity of this assumption and assess the contribution of different sources of noise in establishing the steady state distribution of the pluripotent state, we carry out individual-based simulations for the intermediate switching regime of the network, Fig~\ref{fig:IBM}. In the individual-based model all reactions are treated stochastically thereby accounting for all of the sources of noise in the system. 
The resulting gene expression profiles follow closely those obtained by PDMP simulations, showing that the noise arising from stochastic switching events of promoter configuration accounts for the significant part of overall variability in the network. 
Trajectories of individual transcription factors show that indeed most of the variance in the the molecular distributions are generated by genetic switching events which appear as abrupt stochastic jumps. 
Consistent with experiments, we also find that under LIF gene expression is more heterogeneous than under 2i, but also that the overall levels of expression are higher~\cite{kumar2014deconstructing, singer2014dynamic}. 
In all three regimes of genetic switching supporting pluripotency (LIF+2i,LIF+PD, LIF+CH, LIF and 2i), core transcriptions factors such as Nanog, Oct4, Sox2 are highly expressed. 
The same factors are also repressed in conditions favoring differentiation (CH, PD and none) irrespective of gene switching regime.  This shows that the pluripotent and differentiated states, as determined by the pattern of gene expression, are hardwired in the architecture of the genetic network independent of genetic switching rates. Nevertheless, as we will show in the next section, the routes and dynamics of lineage commitment from pluripotent states strongly depend on the level of molecular noise generated in different gene switching regimes. 


\subsection{Dynamics of lineage commitment is driven by gene switching induced bifurcations of the underlying landscape of gene expression.}

We next ask how the steady state gene expression patterns displayed by the gene network respond to extracellular perturbations in the form of initiation or termination of pluripotency signals. 
Both dual inhibitor 2i (PD+CH) and Leukemia factor LIF based signaling have been shown to provide a stable environment for maintaining pluripotency of stem cells in vitro \cite{martello2014nature,dunn2014defining}. 
Conversely, withdrawal of either LIF or 2i leads to irreversible lineage commitment after a 24-hour period. 
Despite a similar ability to guard pluripotent cells against lineage commitment, these factors deploy different regulatory mechanisms reflected in distinct distributions of pluripotency factors. As a result, stem cell differentiation by withdrawal of different signals proceeds via different routes.
To gain a mechanistic understanding of how the interplay of signaling, molecular noise and network architecture give rise to the steady state expression profiles, we study the dynamics of transitioning between the pluripotent and lineage committed states induced by rapid initiation and withdrawal of signaling conditions (LIF, CH, PD). 

\begin{figure*}[t]
\centering
\includegraphics[width=0.98\textwidth]{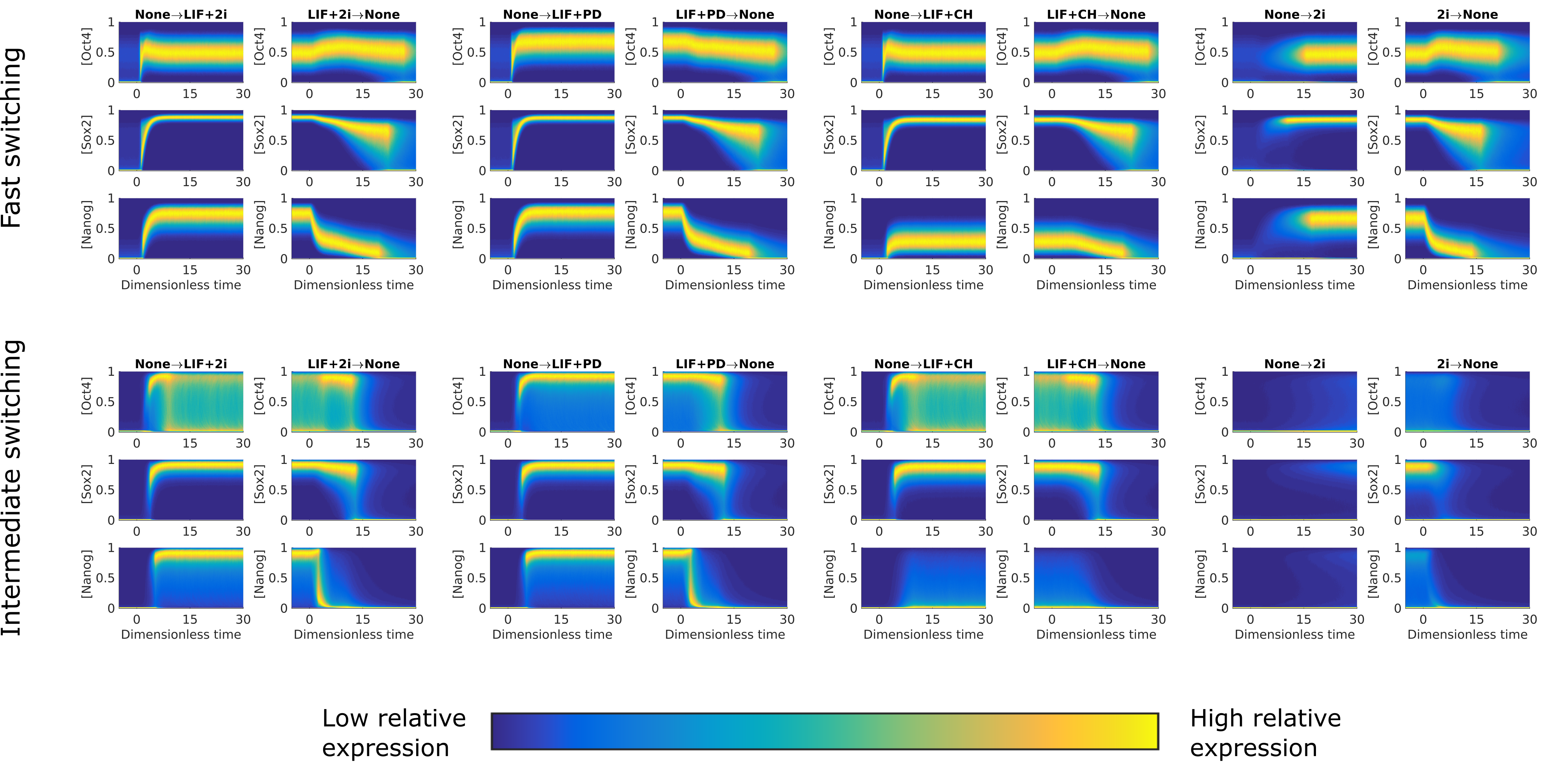}
 \caption{The dynamical behavior of the distributions of transcription factor densities for the intermediate and fast switching regimes. At time $t=0$, the external inputs are changed. The plots show the evolution in probability density for a large ensemble ($10^5$) of PDMP sample paths.}
 \label{fig:dynamical transition}
\end{figure*}

The temporal evolution of distributions of the TFs exiting (LIF/2i withdrawal) and entering (LIF/2i immersion) pluripotent states reveals rich dynamical signatures of these transitions (Fig~\ref{fig:dynamical transition}). 
To reveal the role of stochasticity in these transitions, we compare the intermediate regime---which is dominated by genetic switching---to the fast regime---in which transitions are largely governed by the network topology and the fluctuation of promoter configuration is almost-completely ignored. 
The irreversible nature of transitions manifests clearly in different routes exiting and entering the pluripotent state (transitions to and from the None state in Fig~\ref{fig:dynamical transition}). 

\begin{figure*}[t]
\centering
\includegraphics[width=0.98\textwidth]{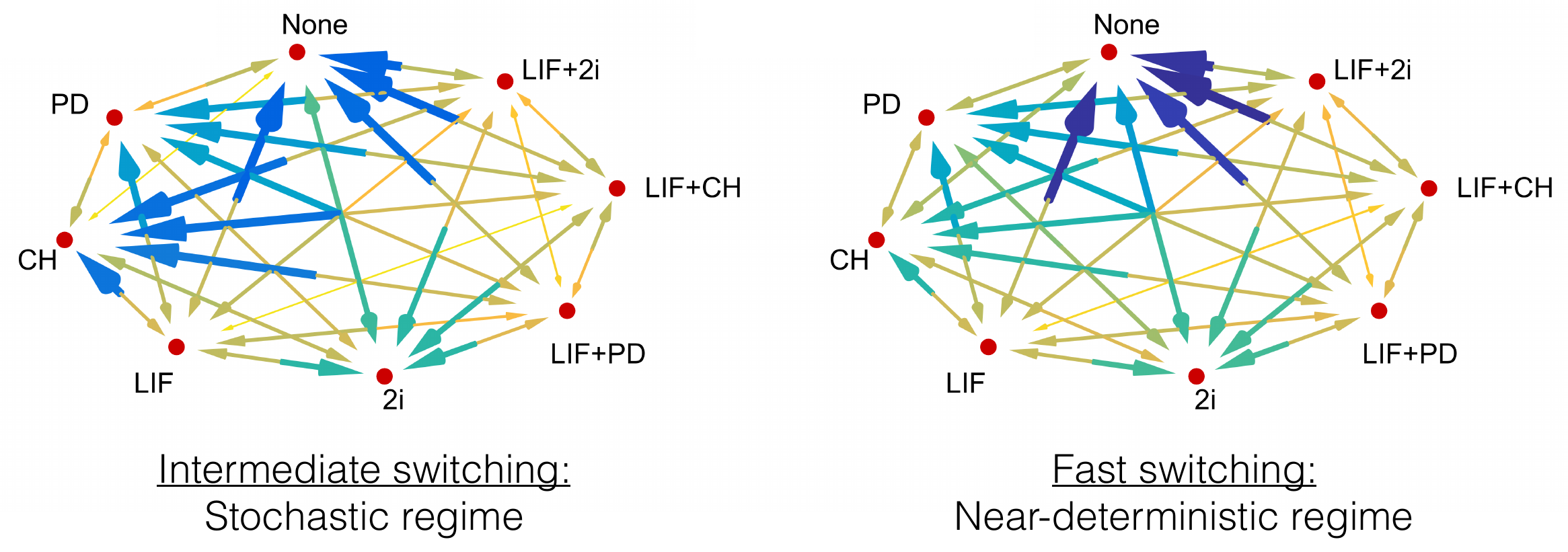}
 \caption{Calculations of transition times between the stationary distributions of different external conditions. A larger, darker arrow indicates that a given transition takes a longer time to converge to its stationary state. This time-scale is measured by simulating a large ensemble ($10^5$) of PDMP sample paths to provide a simulated probability density, and finding the Jensen--Shannon divergence \cite{lin1991divergence,endres2003new} between the instantaneous distribution of each TF and its final stationary distribution. The time for the each divergence to fall below a threshold ($:=0.3$) is recorded, and we choose the largest of these as a quantification of the transition time.}
\label{fig:transition time}
\end{figure*}

In the intermediate regime of genetic switching rates, expression noise greatly facilitates transitions out of a pluripotent state by making the network more responsive to changes in environmental signaling. 
In contrast, in the fast switching regime, upon withdrawal of pluripotency signals the downstream regulation happens on a much slower time-scale with some factors remaining virtually unresponsive to changes of signaling. 
This signaling enhancement in the intermediate regime reveals the importance of molecular noise in making pluripotent states more sensitive to environmental conditions. 
There are qualitatively different patterns of re-entrance into pluripotent states upon LIF vs 2i addition, with LIF being much more efficient at reversing pluripotency compared to the 2i. 
The different potential of signaling culture for pluripotency reversal has been established in experiments \cite{martello2014nature} which have shown that in the later stages of commitment only the LIF is able to reverse lineage-primed cells back to their naive pluripotent states. 
The exit and re-entrance from pluripotency upon withdrawal/addition of LIF shows complex signatures of hysteresis and bifurcations.
This suggests that there can be multiple pathways of entering or exiting pluripotency. 
The transition times for all signaling-induced changes of the steady states of the network are visualized on a kinetic diagram (Fig~\ref{fig:transition time}). 
The kinetic diagram shows an underlying structure to these transitions where conversion among pluripotent states takes place with lower ``activation barriers" compared to transitions accompanying loss of pluripotency.  In the intermediate switching regime there is a clear time-scale separation between transitions which keep cells in pluripotent states and transitions out of pluripotent states.  One may argue that such a time-scale separation between signals inducing differentiation and pluripotency allows embryonic cells to execute developmental decisions more faithfully. In the fast switching regime this clear time-scale separation is partially lost where only the loss of all three signals is separated from the rest of the transitions. 

Detailed analysis of individual distributions and trajectories of transcription factors can be very informative due to their high information content. It is, however, not immediately clear how changes in the expression of individual genes contribute to global changes corresponding to different phenotypic transitions. To reveal such global changes, we project stochastic trajectories of all transcription factors onto the first two eigenvectors obtained by principal component analysis of the reference pluripotent steady state (LIF+2i). Most of the variance of transcription factors is well captured by the few principal components. The high-dimensional steady state of the cellular network can thus be conveniently projected onto a 2-dimensional subspace, allowing us to visualize the attractor states of the network as probability landscapes $\pi(PC_1,PC_2)$ which are often masked by heterogeneous distributions. 

\begin{figure*}[t]
\centering
\includegraphics[width=0.8\textwidth]{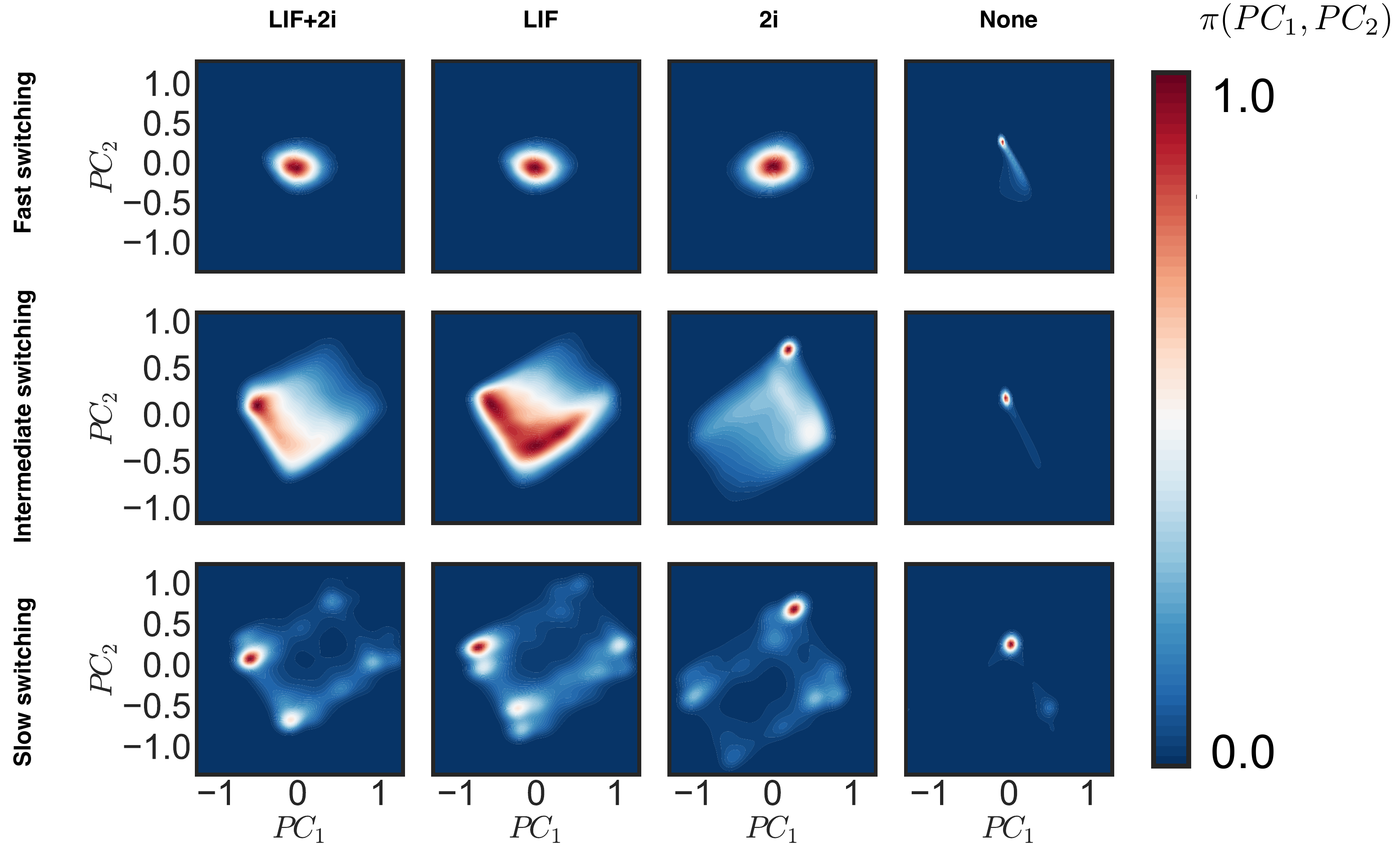}
 \caption{Mapping the cellular attractors of genetic network under different switching and signaling conditions by projecting PDMP simulated gene expression onto first two principal components. The reference state for principal components was chosen to be the LIF+2i/intermediate switching. }
 \label{fig:PCA}
\end{figure*}

Comparing these probability landscapes with different gene switching regimes reveals the distinct roles played by gene switching-induced molecular noise and the deterministic network topology in guiding the transition out of the pluripotent states (Fig~\ref{fig:PCA}). The intermediate gene switching regime, once again, appears to be the more viable regime underlying pluripotent states since the probability landscape shows up as a broad attractor with interconnected states.  Going towards the limit of slow switching results in the fragmentation of the landscape into states separated by high barriers.  This gene switching-induced remodeling of attractors shows the potential for regulation via global epigenetic changes which are purported to act via silencing or activating entire sets of genes at once. Thus one may view gene switching rates as a proxy for genome wide acetylation/methylation patterns which can dramatically alter the access of transcription factors to key target genomic sites. The sequential removal of pluripotency inducing signals reveals a consistent change in the size of the attractor towards occupying smaller regions on the landscape.  This argues for the physical state of network corresponding to pluripotent states to be the one with maximal variance of regulatory transcription factors where lineage commitment is accompanied by their gradual constraining and repression. 
A similar idea which views pluripotency as a macrostate emerging from an ensemble of cells which try to maximizes the information entropy with respect to regulatory transcription factors has been postulated before \cite{macarthur2013statistical,ridden2015entropy}. The analysis of steady state stochastic dynamics of pluripotency network in this work appears in agreement with his view. Furthermore, we are able to reveal the microscopic origin of this entropic paradigm. By analyzing pairwise correlation among different transcription factors we find that signals like LIF/2i create greater independence between the expression of core transcription factors leading them to explore larger range of values. Hence, removal of these signals leads to more constrained and inter-dependent patterns of gene expression for the  same transcription factors which greatly diminishes overall variance. (Fig S1).  
 
\section{Discussion}


Recent studies of ESC have increasingly emphasized systems level perspective on pluripotency and fate determination as outcomes of complex biological computations orchestrated by non-random networks of genes \cite{macarthur2009systems,kontogeorgaki2016noise}. 
Rapid growth of gene expression data collected from controllable {\it in vitro} experiments on mESCs has made it possible to make reliable inferences of gene regulatory networks which govern the state of pluripotency  \cite{dunn2014defining}. These inferences show a highly interconnected nature of regulatory networks where signaling molecules, genomic binding, epigenetics and biochemical feedback act in a concerted manner in balancing pluripotency and decision making. Single-cell experiments have also revealed significant dynamic heterogeneity of gene expression in the population of cells \cite{singer2014dynamic, kumar2014deconstructing}, suggesting the important roles played by molecular noise and non-equilibrium processes. Schematic network topology models are no longer sufficient for rationalizing the level of detailed quantitative information obtained in single-cell studies. 
In the present work we have developed a multi-scale computational scheme for converting experimentally-inferred boolean topologies into quantitative and predictive models of networks with microscopic resolution of gene expression dynamics. The employed computational model is based on previously proposed hybrid stochastic approaches \cite{potoyan2015dichotomous, lin2016bursting, lin2016gene, hufton2016intrinsic} in which the switching dynamics of individual genes are considered exactly while the rest of the biochemical reactions are approximated as deterministic. 
This hybrid stochastic approach is approximately a thousand fold faster than conventional kinetic Monte Carlo methods. 
This allows us to simulate large scale gene regulatory networks of ESC under different culture conditions and gene switching regimes. 
To infer the parameters in the model from the experimental data, we use hybrid simulations to exhaustively sample the space of rates and identify the parameter regime in which the model prediction best matches experimental data. The approximation using the hybrid scheme is validated by carrying out fully stochastic simulations of the network for the identified parameter sets.  This agreement also shows that the switching events of genes---due to stochastic TFs binding to the promoter sites---to be a significant source of variance in the ESC networks. Furthermore we see the changes in gene switching rates as a proxy for global epigenetic modifications that can alter the rates of access of transcription factors to sites buried under chromatin structures. Thus the significant remodeling of steady state landscape that we see by varying the global gene switching rates gives us a glimpse of potentially powerful leverage that epigenetic changes of chromatin have over stability of pluripotent states.  

We find that the intermediate regime, in which gene switching rate is comparable to the other reaction rates in the network, is most consistent with single-cell measurements \cite{singer2014dynamic, kumar2014deconstructing,canham2010functional}. In this regime transcription factors show bursty dynamics which lead to heterogeneous distributions with some showing long tailed and bimodal features. 
We find signaling by LIF and 2i to be a major driving force maintaining the stability of pluripotent states.  Withdrawal of either LIF/2i initiates lineage commitment via a robust pattern of reduced expression of Nanog/Oct4/Sox2 triad. 
To characterize the dynamics of lineage commitment, we compute transition times from pluripotent to differentiated steady states.  We find that higher levels of molecular noise generated by slower gene switching make network more responsive to changes in signaling condition. Next, by carrying out principal component analysis on ensembles of gene expression profiles, we find a much simpler description of pluripotency and lineage commitment in terms of effective probability landscapes.  As the signals safeguarding pluripotency are removed, these landscapes reveal a gradual narrowing of the  steady state attractor explored by the network.  Thus, we see a hierarchical organization of differentiation landscapes where pluripotent states pose the largest attractor which is maintained through the extracellular signals and the molecular noise of gene switching. 

At last we believe the computational framework developed here should also be useful in organize and interpreting experimental data of other complex gene regulatory networks within a coherent and unified computational models thereby making it easier to conceive and vet physical hypothesis in a systematic way.  Given the rapid rise of information from high throughput single-cell nucleic acid based techniques (RNA-seq, RNA-FISH, qPCR, etc), we expect such microscopic resolution models to play important roles for bridging the systems-level behavior of genetic networks with the underlying molecular-level processes of binding and regulation at the genomic sites.

\bibliography{refs}
\end{document}